\documentclass[twocolumn,showpacs]{revtex4}
\usepackage{epstopdf}
\usepackage{graphicx}
\usepackage{color}
\usepackage{dcolumn}
\usepackage{bm}

\def\gappeq{\mathrel{ \rlap{\raise.5ex\hbox{$>$}}
                      {\lower.5ex\hbox{$\sim$}} } }
\def\lappeq{\mathrel{ \rlap{\raise.5ex\hbox{$<$}}
                      {\lower.5ex\hbox{$\sim$}} } }
\newcommand{\ltsimeq}{\raisebox{-0.6ex}{$\,\stackrel
        {\raisebox{-.2ex}{$\textstyle <$}}{\sim}\,$}}
\newcommand{\gtsimeq}{\raisebox{-0.6ex}{$\,\stackrel
        {\raisebox{-.2ex}{$\textstyle >$}}{\sim}\,$}}

\newcommand{\edd}{\varepsilon_{dd}}

\begin{document}

\preprint{PRA}

\title{Exact solutions and stability of rotating dipolar Bose-Einstein condensates in the Thomas-Fermi limit}

\author{R. M. W. van Bijnen$^{1,2,3}$, A. J. Dow$^{1}$, D. H. J. O'Dell$^3$, N. G. Parker$^{3,4}$ and A. M. Martin$^1$}

\address{$^{1}$ School of Physics, University of Melbourne, Parkville,
Victoria 3010, Australia. \\ $^{2}$ Eindhoven University of
Technology, PO Box 513, 5600 MB Eindhoven, The Netherlands. \\$^{3}$
Department of Physics and Astronomy, McMaster University, Hamilton,
Ontario, L8S 4M1, Canada. \\$^4$ School of Food Science and
Nutrition, University of Leeds, LS2 9JT, United Kingdom.}

\date{\today}

\begin{abstract}
We present a theoretical analysis of dilute gas Bose-Einstein
condensates with dipolar atomic interactions under rotation in
elliptical traps. Working in the Thomas-Fermi limit, we employ the
classical hydrodynamic equations to first derive the rotating
condensate solutions and then consider their response to
perturbations. We thereby map out the regimes of stability and
instability for rotating dipolar Bose-Einstein condensates and in
the latter case, discuss the possibility of vortex lattice
formation. We employ our results to propose several novel routes to induce vortex lattice formation in a dipolar
condensate.
\end{abstract}

 \maketitle

\section{Introduction}
The successful Bose-Einstein condensation of $^{52}$Cr atoms
\cite{Griesmaier05,Lahaye07,Koch08} realizes for the first time
Bose-Einstein condensates (BECs) with significant dipole-dipole
interactions. These long-range and anisotropic interactions
introduce rich physical effects, as well as new
opportunities to control BECs. A basic example is how dipole-dipole
interactions modify the shape of a trapped BEC. In a 
prolate (elongated) dipolar gas with the dipoles polarised along
the long axis the net dipolar interaction is attractive, whereas
for an oblate (flattened) configuration with the dipoles aligned
along the short axis the net dipolar interaction is repulsive. As a
result, in comparison to {\it s}-wave BECs (which we define as systems in which atom-atom scattering is dominated by the {\it s}-wave channel), a dipolar BEC
elongates along the direction of an applied polarizing field
\cite{Santos00,Yi01}.  

A full theoretical treatment of a trapped BEC involves solving the Gross-Pitaevskii equation (GPE) for the condensate wave function \cite{Pethick&Smithbook,Pitaevskii&StringariBook}. The non-local nature of the mean-field potential describing dipole-dipole interactions means that this task is significantly harder for dipolar BECs than for {\it s}-wave ones. However, in the limit where the BEC contains a large number of atoms 
the problem of finding the ground state density profile and low-energy dynamics simplifies. In a harmonic trap with oscillator length $a_{\mathrm{ho}}=\sqrt{\hbar/(m \omega)}$, a BEC containing $N$ atoms of mass $m$ which have repulsive {\it s}-wave interactions characterized by scattering length $a$ enters the Thomas-Fermi (TF) regime for large values of the parameter $N a/ a_{\mathrm{ho}}$  \cite{Pethick&Smithbook,Pitaevskii&StringariBook}. In the TF regime the zero-point kinetic energy can be ignored in comparison to the interaction and trapping energies and the Gross-Pitaevskii equation reduces to the equations of superfluid hydrodynamics at $T=0$
\cite{Pethick&Smithbook,Pitaevskii&StringariBook,Pines&NozieresBook}. When applied to a trapped $s$-wave BEC these equations are known to admit a large class of exact analytic solutions  \cite{stringari96}.
The
TF approximation can also be applied to dipolar BECs
\cite{Parker08}. Although the resulting superfluid hydrodynamic equations for a dipolar BEC contain the non-local dipolar potential, exact solutions can still be found \cite{ODell04, Eberlein05} and we make extensive use of them here.
The calculations in this paper are all made within the TF regime.


Condensates are quantum fluids described by a macroscopic wave function $\psi=\sqrt{\rho}\exp[\mathrm{i}S]$, where $\rho$ is the condensate density and $S$ is the condensate phase. This constrains the velocity field $\underline{v}=(\hbar/m) \underline{\nabla} S$
to be curl-free $\underline{\nabla} \times \underline{v}=0$. In an experiment rotation of the  condensate can be accomplished by applying a rotating elliptical deformation to the trapping potential \cite{Madison00,Hodby02}. At low rotation frequencies the elliptical deformation  excites  low-lying collective modes (quadrupole etc.) with quantized angular momentum which may be viewed as surface waves (and which obey $\underline{\nabla} \times \underline{v}=0$). Above a certain critical rotation frequency vortices  are seen to enter the condensate and these satisfy the $\underline{\nabla} \times \underline{v}=0$ condition by having quantized circulation.
 The
hydrodynamic equations for  a BEC provide a simple and accurate description of the low-lying collective modes. Furthermore, they
predict these modes become unstable for certain ranges of rotation frequency \cite{Recati01,Sinha01}. Comparison with experiments
\cite{Madison00,Hodby02} and full numerical simulations of the GPE
\cite{Lundh03,Parker06,Corro07} have clearly shown that the
instabilities are the first step in the entry of vortices into the condensate and the formation of a vortex lattice.
Crucially, the hydrodynamic equations give a clear explanation of
why vortex lattice formation in $s$-wave BECs was only observed to occur at a much
greater rotation frequency than that at which they become energetically favorable.   It is only at these higher frequencies that the
vortex-free condensate becomes dynamically unstable.

Individual vortices \cite{Pu06, ODell07,Wilson08} and vortex
lattices \cite{Cooper05, Zhang05, Cooper07} in dipolar condensates
have already been studied theoretically. However, a key question
that remains is how to make such states in the first place.  In this
paper we extend the TF approximation for rotating trapped
condensates to include dipolar interactions, building on our previous work
\cite{Bijnen07,Martin08}. Specifically, starting from the
hydrodynamic equations of motion we obtain the stationary solutions
for a condensate in a rotating elliptical trap and find when they
become dynamically unstable to perturbations. This enables us to
predict the regimes of stable and unstable motion of a rotating
dipolar condensate. For a non-dipolar BEC (in the TF limit)
the transition between stable and unstable motion is independent of the interaction strength, and depends only
on the rotation frequency and trap ellipticity in the plane perpendicular to the rotation vector
\cite{Recati01,Sinha01}. We show that for a dipolar BEC it is
additionally dependent on the strength of the dipolar interactions
and also the axial trapping strength. All of these
quantities are experimentally tunable and this extends the routes
that can be employed to induce instability. Meanwhile, the critical
rotation frequency at which vortices become energetically favorable $\Omega_v$ is
also sensitive to the trap geometry and dipolar interactions
\cite{ODell07}, and means that the formation of a vortex lattice
following the instability cannot be assumed. Using a simple
prediction for this frequency, we indicate the regimes in which we
expect vortex lattice formation to occur. By considering all of the
key and experimentally tunable quantities in the system we outline
several accessible routes to generate instability and vortex
lattices in dipolar condensates.

This paper is structured as follows. In Section II we introduce the
mean-field theory and the TF approximation for dipolar
BECs, in Section III we derive the hydrodynamic equations for a
trapped dipolar BEC in the rotating frame, and in Section IV we
obtain the corresponding stationary states and discuss their
behaviour. In Section V we show how to obtain the dynamical
stability of these states to perturbations, and in Section VI we
employ the results of the previous sections to discuss possible
pathways to induce instability in the motion of the BEC and discuss
the possibility that such instability leads to the formation of a
vortex lattice. Finally in Section VII we conclude our findings and
suggest directions for future work.

\section{Mean-field theory of a dipolar BEC}
We consider a BEC with long-range dipolar atomic interactions, with
the dipoles aligned in the $z$ direction by an external field. The condensate wave function (mean-field order
parameter) for the condensate $\psi \equiv \psi(\underline{r},t)$
satisfies the GPE which is given by
\cite{Goral00,Santos00,Yi00},
\begin{eqnarray}
i \hbar \frac{\partial \psi}{\partial t}= \left[ -\frac{\hbar^2}{2m}
\nabla^2 + V(\underline{r},t)+\Phi_{dd}(\underline{r},t)+g
\left|\psi \right|^2 \right]\psi
\label{GPE}, \nonumber \\
\end{eqnarray}
where $m$ is the atomic mass.  The $\nabla^2$-term
arises from kinetic energy and $V(\underline{r},t)$ is the external
confining potential. BECs typically feature {\it s}-wave atomic
interactions which gives rise to a local cubic nonlinearity with
coefficient $g=4 \pi \hbar^2 a/m$, where $a$ is the {\it s}-wave
scattering length. Note that $a$, and therefore $g$, can be
experimentally tuned between positive values (repulsive
interactions) and negative values (attractive interactions) by means
of a Feshbach resonance \cite{Lahaye07,Koch08}. The dipolar
interactions lead to a non-local mean-field potential
$\Phi_{dd}(\underline{r},t)$ which is given by \cite{Yi01},
\begin{equation}
\Phi_{dd} (\underline{r},t) = \int d^3 r~U_{dd} \left(
\underline{r}-\underline{r}^{\prime} \right)
\rho\left(\underline{r}^{\prime},t \right), \label{dipole_a}
\end{equation}
where $\rho(\underline{r},t)=|\psi(\underline{r},t)|^2$ is the
condensate density and
\begin{eqnarray}
U_{dd}(\underline{r})=\frac{C_{dd}}{4 \pi}\frac{1-3 \cos^2
\theta}{|\underline{r}|^3} \label{dipole_potential}
\end{eqnarray}
is the interaction potential of two dipoles separated by a vector
$\underline{r}$, where $\theta$ is the angle between $\underline{r}$
and the polarization direction, which we take to be the $z$-axis. The dipolar BECs made to date have featured permanent
magnetic dipoles. Then, assuming the dipoles to have moment $d_m$
and be aligned in an external magnetic field
$\underline{B}=\hat{k}B$, the dipolar coupling is $C_{dd}= \mu_0
d_m^2$ \cite{Goral00}, where $\mu_0$ is the permeability of free
space. Alternatively, for dipoles induced by a static electric field
$\underline{E}=\hat{k}E$, the coupling constant $C_{dd}=E^2
\alpha^2/\epsilon_0$ \cite{Yi00,You98}, where $\alpha$ is the static
polarizability and $\epsilon_0$ is the permittivity of free space.
In both cases, the sign and magnitude of $C_{dd}$ can be tuned
through the application of a fast-rotating external field
\cite{Giovanazzi02}.

We will specify the interaction strengths through the parameter
\begin{equation}
\varepsilon_{dd}=\frac{C_{dd}}{3g},
\end{equation}
which is the ratio of the dipolar interactions to the s-wave
interactions \cite{Giovanazzi02}. We take the {\it s}-wave
interactions to be repulsive, $g>0$, and so where we discuss
negative values of $\varepsilon_{dd}$, this corresponds to
$C_{dd}<0$. We will also limit our analysis to the regime of $-0.5 <
\varepsilon_{dd} < 1$, where the Thomas-Fermi approach predicts that
non-rotating stationary solutions are robustly stable
\cite{ODell04}. Outside of this regime the situation
becomes more complicated since the non-rotating system becomes prone
to collapse \cite{Parker09}.

We are concerned with a BEC confined by an elliptical harmonic trapping potential
of the form,
\begin{equation}\label{harmonicV}
V(\underline{r}) = \frac{1}{2} m \omega_{\perp}^2 \left[(1 -
\epsilon) x^2 + (1 + \epsilon)y ^2 + \gamma^2 z^2\right].
\end{equation}
In the $x-y$ plane the trap has mean trap frequency $\omega_{\perp}$
and ellipticity $\epsilon$. The trap strength in the axial
direction, and indeed the geometry of the trap itself, is specified
by the trap ratio $\gamma=\omega_z/\omega_{\perp}$. When $\gamma \gg
1$ the BEC shape will typically be oblate (flattened) while for
$\gamma \ll 1$ it will typically be prolate (elongated), although for strong enough dipolar interactions the electrostrictive/magnetostrictive effect can cause a BEC in an oblate trap to become prolate itself.

The time-dependent GPE (\ref{GPE}) can be reduced to its
time-independent form by making the substitution
$\psi(\underline{r},t)=\sqrt{\rho(\underline{r})}\exp(i\mu
t/\hbar)$, where $\mu$ is the chemical potential of the system.
We employ the TF approximation whereby the kinetic
energy of static solutions is taken to be negligible in comparison
to the potential and interaction energies.  The validity of this
approximation in dipolar BECs has been discussed elsewhere
\cite{Parker08}. Then, the time-independent GPE
reduces to,
\begin{equation}\label{time_indep_GPE_harmonicV}
V(\underline{r}) + \Phi_{dd}(\underline{r})+ g \rho(\underline{r}) =
\mu.
\end{equation}
 For ease of
calculation the dipolar potential $\Phi_{dd}(\underline{r}) $ can be
expressed as,
\begin{equation}
\Phi_{dd} (\underline{r}) = -3g \varepsilon_{dd}
\left(\frac{\partial^2}{\partial z^2} \phi(\underline{r})
+\frac{1}{3} \rho(\underline{r}) \right), \label{eq:Phidd}
\label{dipole}
\end{equation}
where $\phi(\underline{r})$ is a fictitious `electrostatic'
potential defined by \cite{ODell04,Eberlein05},
\begin{equation}
\phi(\underline{r}) = \frac{1}{4 \pi} \int \frac{d^3 r^{\prime}
\rho(\underline{r}^{\prime})}{\left|
\underline{r}-\underline{r}^{\prime} \right|}. \label{potential}
\end{equation}
This effectively reduces the problem of calculating the dipolar
potential (\ref{dipole_a}) to the calculation of an electrostatic potential of
the form (\ref{potential}), for which a much larger theoretical body of literature
exists. Exact solutions of Eq.\ (6) for $\rho(\underline{r})$,
$\phi(\underline{r})$ and hence $\Phi_{dd}(\underline{r})$ can be
obtained for any general parabolic trap, as proven in Appendix A of
Ref. \cite{Eberlein05}. In particular, the solutions of
$\rho(\underline{r})$ take the form
\begin{eqnarray}
\rho(\underline{r})=\rho_0\left(1-\frac{x^2}{R_x^2}-\frac{y^2}{R_y^2}-\frac{z^2}{R_z^2}\right)
\,\,\,\, {\rm for} \,\,\, \rho(\underline{r}) \ge 0 \label{TF}
\end{eqnarray}
where $\rho_0=15N/(8 \pi R_x R_y R_z)$ is the central density.
Remarkably, this is the general inverted parabola density profile
familiar from the TF limit of non-dipolar BECs. An important
distinction, however, is that for the dipolar BEC the aspect ratio
of the parabolic solution differs from the trap aspect ratio.

\section{Hydrodynamic Equations in the Rotating Frame}
Having introduced the TF model of a dipolar BEC we now extend this
to include rotation and derive hydrodynamic equations for the
rotating system. We consider the rotation to act about the $z$-axis,
described by the rotation vector $\underline{\Omega}$ where
$\Omega=|\underline{\Omega}|$ is the rotation frequency and the
Hamiltonian  in the rotating frame is given by,
\begin{equation}\label{Heff}
H_{{\rm eff}} = H_0 - \underline{\Omega} \cdot \hat{L},
\end{equation}
where $H_0$ is the Hamiltonian in absence of the rotation and
$\hat{L} = -i\hbar (\underline{r} \times \underline{\nabla})$ is the quantum
mechanical angular momentum operator. Using this result with the
Hamiltonian $H_0$ from Eq.\ (\ref{GPE}) we obtain
\cite{Leggett_Book,Leggett00},
\begin{eqnarray}
i \hbar \frac{\partial \Psi(\underline{r},t)}{ \partial t} &=&
\left[-\frac{\hbar^2}{2m}\nabla^2 + V(\underline{r}) +
\Phi_{dd}(\underline{r},t)+g|\Psi(\underline{r},t) |^2 \right.\nonumber \\
&-& \left.\Omega\frac{\hbar}{i} \left(x \frac{\partial}{\partial y}
- y \frac{\partial}{\partial x} \right)\right]
\Psi(\underline{r},t).\label{GPE_rotating_d}
\end{eqnarray}
Note that all space coordinates $\underline{r}$ are
those of the rotating frame and the time independent trapping
potential $V(\underline{r})$, given by Eq.\ (\ref{harmonicV}), is stationary in this frame.  Momentum coordinates, however, are expressed in the laboratory frame \cite{Leggett_Book,Leggett00,Lifshitz}.

We can express the condensate mean field in terms of a density
$\rho(\underline{r},t)$ and phase $S(\underline{r},t)$ as
$\psi(\underline{r},t)=\sqrt{\rho(\underline{r},t)}\exp[iS(\underline{r},t)]$,
and so that the condensate velocity is $\underline{v} =
(\hbar/m)\underline{\nabla} S$. Substuting into the time-dependent GPE
(\ref{GPE_rotating_d}) and equating imaginary and real terms leads to the
following equations of motion,
\begin{eqnarray}\label{cont_eq_d}
\frac{\partial \rho}{\partial t} &=& -\underline{\nabla} \cdot \left[
\rho\left(\underline{v} - \underline{\Omega} \times
\underline{r}\right) \right],\\
m \frac{\partial \underline{v}}{\partial t} &=& -\underline{\nabla} \left(
\frac{1}{2} m \underline{v} \cdot \underline{v} + V(\underline{r}) +
\Phi_{dd}(\underline{r})
 \right. \nonumber \\
&&+ \left. g \rho - m \underline{v} \cdot \left[ \underline{\Omega}
\times \underline{r} \right] \right).\label{motion_eq_d}
\end{eqnarray}
In the absence of dipolar interactions ($\Phi_{dd}=0$) Eqs.\
(\ref{cont_eq_d}) and (\ref{motion_eq_d}) are commonly known as the
superfluid hydrodynamic equations \cite{Pethick&Smithbook,Pitaevskii&StringariBook,Pines&NozieresBook} since
they resemble the equation of continuity and the Euler equation of
motion from dissipationless fluid dynamics. Here we have extended them to
include dipolar interactions.

Note that the form of condensate velocity leads to the
relation,
\begin{equation}\label{irrotationality}
\underline{\nabla} \times \underline{v} = \frac{\hbar}{m}\underline{\nabla} \times \underline{\nabla} S=
0,
\end{equation}
which immediately reveals that the condensate is irrotational. The
exceptional case is when the velocity potential $(\hbar/m)S$ is
singular, which arises when a quantized vortex occurs in the system.


\section{Stationary Solution of the Hydrodynamic Equations}\label{SecStationaryStates}
We now search for stationary solutions of the hydrodynamic Eqs.\
(\ref{cont_eq_d}) and (\ref{motion_eq_d}). These states satisfy the
equilibrium conditions,
\begin{equation}\label{stationary_solutions}
\frac{\partial \rho}{\partial t} = 0, \hspace{2cm} \frac{\partial
\underline{v}}{\partial t} = 0.
\end{equation}
Following the approach of Recati {\it et al}. \cite{Recati01} we
assume the velocity field ansatz,
\begin{equation}\label{velocity_ansatz}
\underline{v} = \alpha \underline{\nabla}(x y).
\end{equation}
Here $\alpha$ is a velocity field amplitude that will provide us
with a key parameter to parameterise our rotating solutions. Note
that this is the velocity field in the laboratory frame expressed in
the coordinates of the rotating frame, and also that it satisfies
the irrotationality condition (\ref{irrotationality}). Combining
Eqs.\ (\ref{motion_eq_d}) and (\ref{velocity_ansatz}) we obtain the
relation,
\begin{eqnarray}
\mu=\frac{m}{2} \left(\tilde{\omega}_x^2 x^2 +\tilde{\omega}_y^2 y^2
+\omega_z^2 z^2
\right)+g\rho(\underline{r})+\Phi_{dd}(\underline{r}), \label{mu}
\end{eqnarray}
where the \textit{effective} trap frequencies $\tilde{\omega}_x$ and
$\tilde{\omega}_y$ are given by,
\begin{equation}\label{omax}
\tilde{\omega}_x^2 = \omega_{\perp}^2(1 - \epsilon) + \alpha^2 - 2
\alpha \Omega
\end{equation}
\begin{equation}\label{omay}
\tilde{\omega}_y^2 = \omega_{\perp}^2(1 + \epsilon) + \alpha^2 + 2
\alpha \Omega.
\end{equation}
The dipolar potential inside an inverted parabola density
profile (\ref{TF}) has been found in Refs.
\cite{Eberlein05,Bijnen07} to be,
\begin{eqnarray}
\frac{\Phi_{dd}}{3g\varepsilon_{dd}}&=&\frac{\rho_0 \kappa_x
\kappa_y}{2}\left[\beta_{001}-\frac{x^2\beta_{101}+y^2\beta_{011}+3z^2\beta_{002}}{R_z^2}\right]
- \frac{\rho}{3} \nonumber \\
\end{eqnarray}
where we have defined the condensate aspect ratios
$\kappa_x=R_x/R_z$ and $\kappa_y=R_y/R_z$, and where the
coefficients $\beta_{ijk}$ are given by,
\begin{eqnarray}
\beta_{ijk}=\int_0^{\infty}\frac{ds}{\left(\kappa_x^2+s\right)^{i+\frac{1}{2}}
\left(\kappa_y^2+s\right)^{j+\frac{1}{2}}
\left(1+s\right)^{k+\frac{1}{2}}},
\end{eqnarray}
where $i$, $j$ and $k$ are integers.
Note that for the cylindrically symmetric case, where $\kappa_x =
\kappa_y = \kappa$, the integrals $\beta_{ijk}$ evaluate to
\cite{Gradshteyn00},
\begin{eqnarray}
\beta_{ijk}=2\frac{_2F_1\left(k+\frac{1}{2},1;i+j+k+\frac{3}{2};1-\kappa^2\right)}{\left(1+2i+2j+2k\right)\kappa^{2(i+j)}}
\end{eqnarray}
where $_2F_1$ denotes the Gauss hypergeometric function
\cite{Abramowitz74}.
Thus we can rearrange
Eq.\ (\ref{mu}) to obtain an expression for the density profile,
\begin{eqnarray}
\rho&=&\frac{\mu -\frac{m}{2} \left(\tilde{\omega}_x^2 x^2 +\tilde{\omega}_y^2 y^2 +\omega_z^2 z^2 \right)}{g\left(1-\varepsilon_{dd}\right)} \nonumber \\
&+&\frac{3g\varepsilon_{dd}\frac{n_0\kappa_x \kappa_y}{2R_z^2}\left[x^2\beta_{101}+y^2\beta_{011}+3z^2\beta_{002}-R_z^2\beta_{001}\right] }{g\left(1-\varepsilon_{dd}\right)}. \nonumber \\
\label{rho}
\end{eqnarray}
Comparing the $x^2$, $y^2$ and $z^2$ terms in Eq.~(\ref{TF}) and
Eq.~(\ref{rho}) we find three self-consistency relations that define
the size and shape of the condensate:
\begin{eqnarray}
\kappa_{x}^2&=&\left(\frac{\omega_z}{\tilde{\omega}_{x}}\right)^2
\frac{1+\varepsilon_{dd}\left(\frac{3}{2}\kappa_x^3 \kappa_y
\beta_{101}-1\right)}{\zeta} \label{kx},
\\
\kappa_y^2&=&\left(\frac{\omega_z}{\tilde{\omega}_y}\right)^2 \frac{1+\varepsilon_{dd}\left(\frac{3}{2}\kappa_y^3\kappa_x \beta_{011}-1\right)}{\zeta} \label{ky}, \\
R_z^2&=&\frac{2g\rho_0}{m\omega_z^2}\zeta, \label{size}
\end{eqnarray}
where $\zeta=1-\varepsilon_{dd}\left[1-\frac{9 \kappa_x
\kappa_y}{2}\beta_{002}\right]$. Furthemore, by inserting
 Eq.~(\ref{rho}) into
 Eq.~(\ref{cont_eq_d}) we find that stationary solutions satisfy the condition,
\begin{eqnarray}
0&=&\left(\alpha+\Omega\right)\left(\tilde{\omega}_x^{2}-\frac{3}{2}\varepsilon_{dd} \frac{\omega_{\perp}^2\kappa_x\kappa_y \gamma^{2}}{\zeta} \beta_{101}\right) \nonumber \\
&+&\left(\alpha-\Omega\right)\left(\tilde{\omega}_y^{
2}-\frac{3}{2}\varepsilon_{dd} \frac{\omega_{\perp}^2\kappa_x
\kappa_y \gamma^{2}}{\zeta} \beta_{011}\right). \label{alpha}
\end{eqnarray}
We can now solve Eq.~(\ref{alpha}) to give the velocity field
amplitude $\alpha$ for a given $\varepsilon_{dd}$, $\Omega$ and trap
geometry. In the limit $\varepsilon_{dd}=0$ this amplitude is
independent of the {\it s}-wave interaction strength $g$ and the
trap ratio $\gamma$. However, in the presence of dipolar
interactions the velocity field amplitude becomes dependent on both
$g$ and $\gamma$. For fixed $\varepsilon_{dd}$ and trap geometry,
Eq.~(\ref{alpha}) leads to branches of $\alpha$ as a function of
rotation frequency $\Omega$. These branches are
significantly different between traps that are circular
($\epsilon=0$) or elliptical ($\epsilon>0$) in the $x-y$ plane, and
so we will consider each case in turn. Note that we restrict our
analysis to the range $\Omega<\omega_{\perp}$: for $\Omega \sim
\omega_{\perp}$ the static solutions can disappear, with the
condensate becoming unstable to a centre-of-mass instability
\cite{Recati01}.

\subsection{Circular trapping in the $x-y$ plane: $\epsilon=0$}
\begin{figure}
\centering
\includegraphics[width=7.5cm]{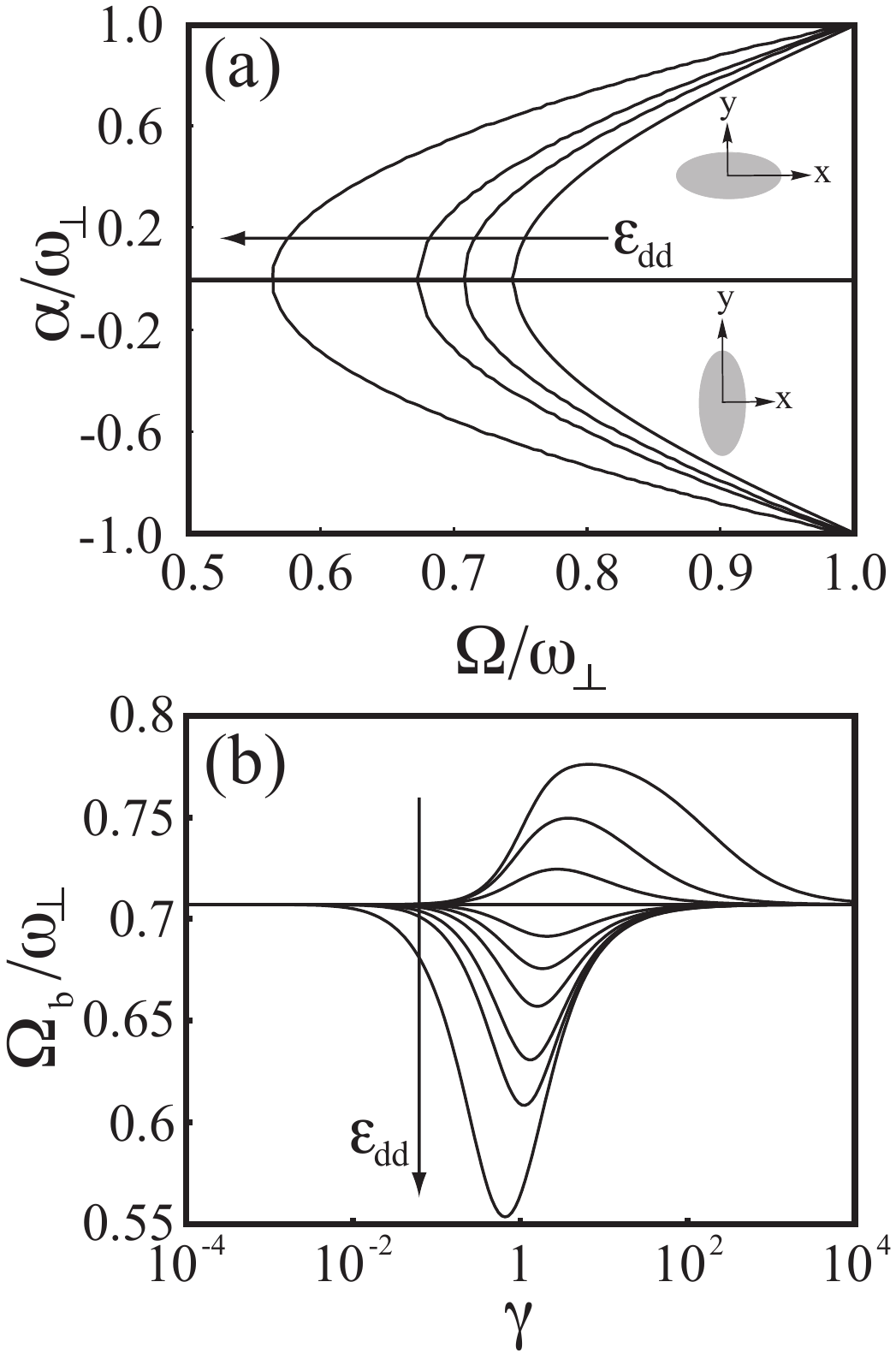}
\caption{(a) Irrotational velocity field amplitude $\alpha$ of the
static condensate solutions as a function of the trap rotation
frequency $\Omega$ in a spherically-symmetric trap ($\gamma=1$ and
$\epsilon=0$). Various values of $\varepsilon_{dd}$ are presented:
$\varepsilon_{dd}=-0.49$, $0$, $0.5$ and $0.99$. Insets illustrate the
geometry of the condensate in the $x-y$ plane. (b) The bifurcation
frequency $\Omega_b$ (the point at which the solutions of $\alpha$
in (a) bifurcate) according to Eq.\ (\ref{eq:Bif}) versus trap ratio
$\gamma$. Plotted are the results for $\varepsilon_{dd}=-0.49$,
$-0.4$, $-0.2$, $0$, $0.2$, $0.4$, $0.6$, $0.8$, $0.9$ and $0.99$.
In (a) and (b) $\varepsilon_{dd}$ increases in the direction of the
arrow. }\label{Fig1}
 \vspace{-0.5cm}
\end{figure}

We first consider the case of a trap with no
ellipticity in the $x-y$ plane ($\epsilon=0$). In Fig.\
\ref{Fig1}(a) we plot the solutions of Eq.~(\ref{alpha}) as a
function of rotation frequency $\Omega$ for a spherically-symmetric
trap $\gamma=1$ and for various values of $\varepsilon_{dd}$. Before
dicussing the specific cases, let us first point out that for each
$\varepsilon_{dd}$ the solutions have the same qualitative
structure. Up to some critical rotation frequency only one solution
exists corresponding to $\alpha=0$. At this critical point the
solution bifurcates, giving two additional solutions for $\alpha>0$
and $\alpha<0$ on top of the original $\alpha=0$ solution. We term
this critical frequency the bifurcation frequency $\Omega_b$.

For $\varepsilon_{dd}=0$ we regain the results of Refs.\
\cite{Recati01,Sinha01} with a bifurcation point at
$\Omega_b=\omega_\perp/\sqrt{2}$ and, for $\Omega > \Omega_b$, non-zero
solutions given by $\alpha=\pm \sqrt{2
\Omega^2-\omega_\perp^2}/\omega_\perp$ \cite{Recati01}. The physical
significance of the bifurcation frequency has been established for
the non-dipolar case and is related to the fact that the system
becomes energetically unstable to the spontaneous excitation of quadrupole modes for $\Omega\geq
\omega_\perp/\sqrt{2}$. In the TF limit, a general surface
excitation with angular momentum $\hbar l= \hbar q_{l} R$, where $R$
is the TF radius and $q_{l}$ is the quantized wave number, obeys the
classical dispersion relation $\omega_{l}^2=(q_{l}/m) \nabla_{R} V$
involving the local harmonic potential $V=m \omega_{\perp}^{2}R^{2}/2$
evaluated at $R$ \cite{Pitaevskii&StringariBook}. Consequently, for
the non-rotating and non-dipolar BEC
$\omega_{l}=\sqrt{l}\omega_\perp$. Meanwhile, inclusion of the
rotational term in the Hamiltonian (\ref{Heff}) shifts the mode
frequency by $-l\Omega$. Then, in the rotating frame, the frequency
of the $l=2$ quadrupole surface excitation becomes
$\omega_2(\Omega)=\sqrt{2}\omega_\perp-2\Omega$
\cite{Pitaevskii&StringariBook}. The bifurcation frequency thus
coincides with the vanishing of the energy of the quadrupolar mode
in the rotating frame, and the two additional solutions arise from
excitation of the quadrupole mode for $\Omega \ge
\omega_\perp/\sqrt{2}$.

For the non-dipolar BEC it is noteworthy that $\Omega_b$ does not
depend on the interactions. This feature arises because the mode
frequencies $\omega_l$ themselves are independent of $g$. However,
in the case of long-range dipolar interactions the potential
$\Phi_{dd}$ of Eq.\ (\ref{eq:Phidd}) gives non-local contributions,
breaking the simple dependence of the force $-\nabla V$ upon $R$
\cite{ODell04}. Thus we expect the resonant condition for exciting
the quadrupolar mode, i.e. $\Omega_b=\omega_l/l$ (with $l=2$), to
change with $\varepsilon_{dd}$. In Fig.\ \ref{Fig1}(a) we see that this is the
case: as dipole interactions are introduced, our solutions change
and the bifurcation point, $\Omega_b$, moves to lower (higher)
frequencies for $\varepsilon_{dd}>0$ ($\varepsilon_{dd}<0$). Note
that the parabolic solution still satisfies the hydrodynamic
equations providing $-0.5<\varepsilon_{dd}<1$. Outside of this range the parabolic solution may still exist but it is no longer guaranteed to be stable against perturbations.

Density profiles for $\alpha=0$ have zero ellipticity in the $x-y$
plane. By contrast, the $|\alpha|>0$ solutions have an elliptical
density profile, even though the trap itself has zero ellipticity.
This remarkable feature arises due to a spontaneous breaking of the
axial rotational symmetry at the bifurcation point. For $\alpha>0$
the condensate is elongated in $x$ while for $\alpha<0$ it is
elongated in $y$, as illustrated in the insets in Fig.~\ref{Fig1}(a). In the
absence of dipolar interactions the $|\alpha|>0$ solutions can be
intepreted solely in terms of the effective trapping frequencies
${\tilde \omega}_x$ and ${\tilde \omega}_y$ given by Eqs.\
(\ref{omax}) and (\ref{omay}). The introduction of dipolar
interactions considerably complicates this picture, since they also
modify the shape of the solutions. Notably, for $\varepsilon_{dd}>0$
the dipolar interactions make the BEC more prolate, i.e., reduce
$\kappa_x$ and $\kappa_y$, while for $\varepsilon_{dd}<0$ they make
the BEC more oblate, i.e., increase $\kappa_x$ and $\kappa_y$.

In Fig.\ \ref{Fig1}(a) we see that as the dipole interactions are increased
the bifurcation point $\Omega_b$ moves to lower frequencies. The
bifurcation point can be calculated analytically as follows. First,
we note that for $\alpha = 0$ the condensate is cylindrically
symmetric and $\kappa_x = \kappa_y = \kappa$. In this case the aspect ratio
$\kappa$ is determined by the transcendental equation \cite{Yi00,ODell04,Eberlein05}
\begin{eqnarray}
\left[\left(\frac{\gamma^2}{2}+1\right)\frac{f(\kappa)}{1-\kappa^2}-1\right]+\frac{\left(\varepsilon_{dd}-1\right)\left(\kappa^2-\gamma^2\right)}{3\kappa^2\varepsilon_{dd}}=0
\nonumber \\
\end{eqnarray}
where
\begin{eqnarray}
f(\kappa)=\frac{2+\kappa^2\left[4-3\beta_{000}\right]}{2\left(1-\kappa^2\right)}
\end{eqnarray}
with $\beta_{000}=(1/\sqrt{1-\kappa^2}) \ln [ (1+\sqrt{1-\kappa^2})/(1-\sqrt{1-\kappa^2})]$ for the prolate case ($\kappa < 1$), and $\beta_{000}=(2/\sqrt{\kappa^2-1}) \arctan [\sqrt{\kappa^2-1}]$ for the oblate case ($\kappa > 1$).
For small $\alpha \rightarrow 0_+$, we can calculate the first order
corrections to $\kappa_x$ and $\kappa_y$ with respect to $\kappa$
from Eqs.\ (\ref{kx},\ref{ky}). We can then insert these values in
Eq.\ (\ref{alpha}) and solve for $\Omega$, noting that in the limit
$\alpha \rightarrow 0$ we have $\Omega \rightarrow \Omega_b$. Thus,
we find
\begin{eqnarray}
\label{Omega_b}
\frac{\Omega_b}{\omega_{\perp}}=\sqrt{\frac{1}{2}+\frac{3}{4}\kappa^2\varepsilon_{dd}\gamma^2\frac{\kappa^2\beta_{201}-\beta_{101}}{1-\varepsilon_{dd}\left(1-\frac{9}{2}\kappa^2\beta_{002}\right)}}
\label{eq:Bif}.
\end{eqnarray}
 In Fig.\ \ref{Fig1}(b) we plot $\Omega_b$ [Eq.\
(\ref{Omega_b})] as a function of $\gamma$ for various values of
$\varepsilon_{dd}$. For $\varepsilon_{dd}=0$ we find that the
bifurcation point remains unaltered at $\Omega_b=\omega_x/\sqrt{2}$
as $\gamma=\omega_z/\omega_x$ is changed \cite{Recati01,Sinha01}. As
$\varepsilon_{dd}$ is increased the value of $\gamma$ for which
$\Omega_b$ is a minimum changes from a trap shape which is oblate
($\gamma
> 1$) to prolate ($\gamma < 1$). Note that for $\varepsilon_{dd}=0.99$ the minimum bifurcation frequency occurs at $\Omega_b \approx 0.55$, which is over a $20\%$ deviation from the non-dipolar value. For more extreme values of $\varepsilon_{dd}$ we can expect $\Omega_b$ to deviate even further, although the validity of the inverted parabola TF solution does not necessary hold. For a fixed $\gamma$ we also find that as $\varepsilon_{dd}$ increases the bifurcation frequency decreases monotonically.
\begin{figure}
\centering
\includegraphics[width=7.5cm]{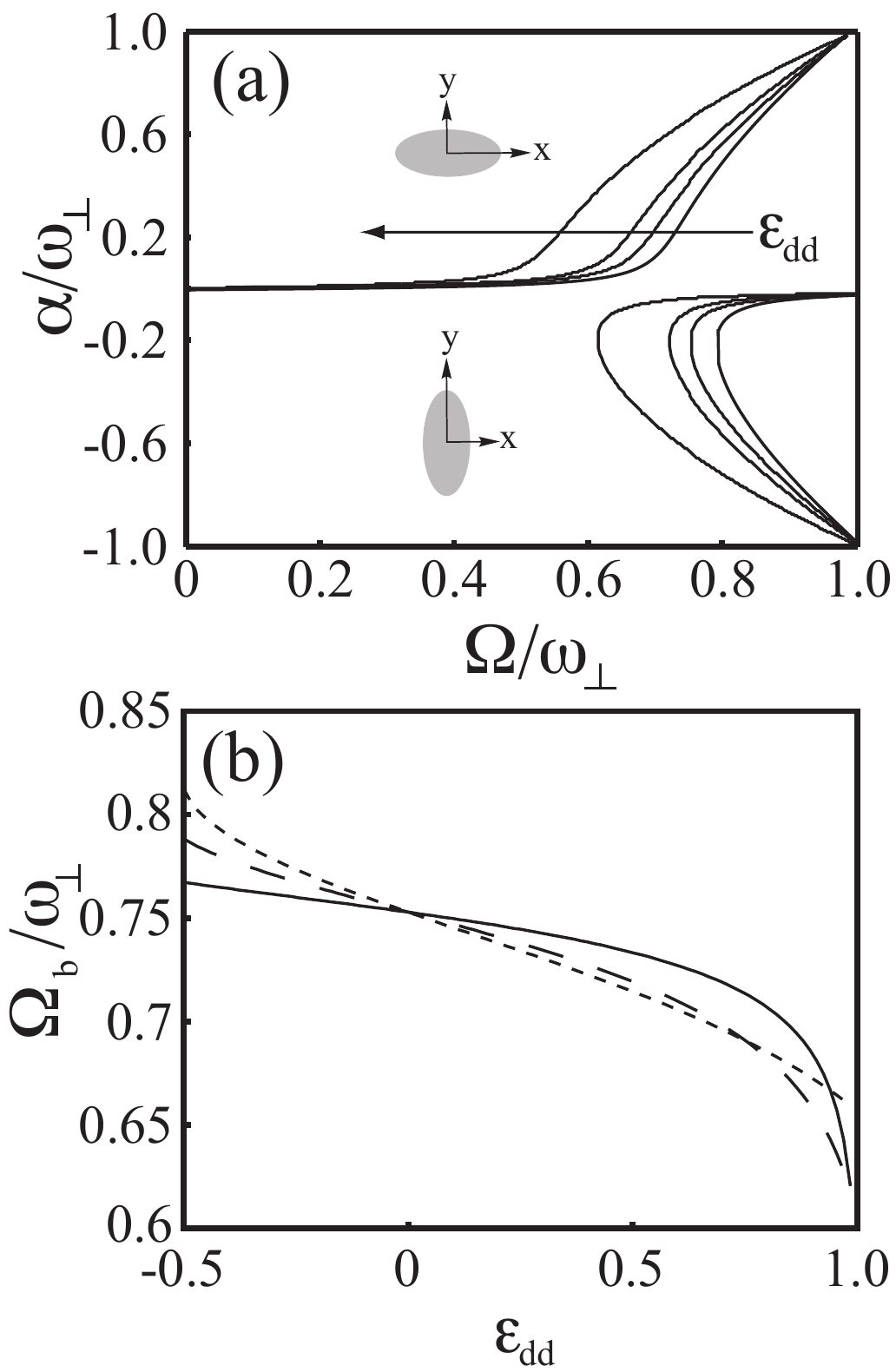}
\caption{ (a) Irrotational velocity field amplitude, $\alpha$, as a
function of the trap rotation frequency, $\Omega$, for a trap ratio
$\gamma=1$ and ellipticity $\epsilon=0.025$. Various values of
$\varepsilon_{dd}$ are presented, $\varepsilon_{dd}=-0.49$, $0$,
$0.5$ and $0.99$, with $\varepsilon_{dd}$ increasing in the
direction of the arrow. Insets illustrate the geometry of the
condensate in the $x-y$ plane. (b) Backbending point $\Omega_b$
versus $\varepsilon_{dd}$ for $\epsilon=0.025$ and $\gamma=0.5$
(solid curve), $1.0$ (long dashed curve) and $2.0$ (short dashed
curve).\label{Fig2}}
 \vspace{-0.5cm}
\end{figure}

\subsection{Elliptical trapping in the $x-y$ plane: $\epsilon > 0$}
Consider now the effect of  finite ellipticity in the
$x-y$ plane ($\epsilon > 0$). Rotating elliptical traps have been
created experimentally with laser and magnetic fields
\cite{Madison00,Hodby02}. Following the experiment of Madison {\it
et al.} \cite{Madison00} we will employ a weak trap ellipticity of
$\epsilon=0.025$. In Fig.\ \ref{Fig2}(a) we have plotted the
solutions to Eq.~(\ref{alpha}) for various values of
$\varepsilon_{dd}$ in a $\gamma=1$ trap. As predicted for
non-dipolar interactions \cite{Recati01,Sinha01} the solutions
become heavily modified for $\epsilon>0$. There exists an upper
branch of $\alpha>0$ solutions which exists over the whole range of
$\Omega$, and a lower branch of $\alpha<0$ solutions which
back-bends and is double-valued. We term the frequency at which the
lower branch back-bends to be the back-bending frequency $\Omega_b$.
The bifurcation frequency in non-elliptical traps can be regarded as
the limiting case of the back-bending frequency, with the differing
nonclamenture employed to emphasise the different structure of the
solutions at this point. However, for convenience we will employ the
same parameter for both, $\Omega_b$. No $\alpha=0$ solution exists
(for any non-zero $\Omega$). In the absence of dipolar interactions
the effect of increasing the trap ellipticity is to {\em increase}
the back-bending frequency $\Omega_b$. Turning on the dipolar
interactions, as in the case of $\epsilon=0$, {\em reduces}
$\Omega_b$ for $\varepsilon_{dd} > 0$, and {\em increases}
$\Omega_b$ for $\varepsilon_{dd} < 0$. This is more clearly seen in
Fig.\ \ref{Fig2}(b) where $\Omega_b$ is plotted versus
$\varepsilon_{dd}$ for various values of the trap ratio $\gamma$.
Also, as in the $\epsilon=0$ case, increasing $\varepsilon_{dd}$
decreases both $\kappa_x$ and $\kappa_y$, i.e. the BEC becomes more
prolate.

Importantly, the back-bending of the lower branch can
introduce an instability. Consider the BEC to be on the lower branch
at some fixed rotation frequency $\Omega$.  Now consider decreasing
$\varepsilon_{dd}$.  The back-bending frequency $\Omega_b$ increases
and at some point can exceed $\Omega$.  In other words, the static
solution of the BEC suddenly disappears and the BEC finds itself in
an unstable state.  We will see in Section VI that this type of
instability can also be induced by variations in $\gamma$ and
$\epsilon$.

\section{Dynamical Stability of Stationary Solutions}\label{SecStability}
Although the solutions derived above are static solutions in the rotating frame they are
not necessarily stable, and so in this section we analyze their
dynamical stability. Consider small perturbations in the BEC density
and phase of the form $\rho=\rho_0+\delta \rho$ and $S=S_0+\delta
S$. Then, by linearizing the hydrodynamic equations
Eqs.~(\ref{cont_eq_d}, \ref{motion_eq_d}), the dynamics of such
perturbations can be described as,
 \begin{eqnarray}
 \label{stability}
\frac{\partial }{\partial t} \left[\begin{array}{c}
 \delta S \\
  \delta \rho \\
\end{array}
\right] = -\left[\begin{array}{cc}
 \underline{v}_c \cdot \nabla & g\left(1+\varepsilon_{dd}K\right)/m \\
 \nabla \cdot \rho_0
\nabla & \left[\left(\nabla \cdot
\underline{v}\right)+\underline{v}_c \cdot \nabla \right] \\
\end{array}
\right] \left[\begin{array}{c}
 \delta S \\
  \delta \rho \\
\end{array}
\right]
 \end{eqnarray}
where
$\underline{v}_c=\underline{v}-\underline{\Omega}\times\underline{r}$
and the integral operator $K$ is defined as
\begin{eqnarray}
(K \delta \rho)(\underline{r})=-3\frac{\partial^2}{\partial
z^2}\int\frac{\delta\rho(\underline{r}^{\prime})
\ d\underline{r}^{\prime}}{4\pi\left|\underline{r}-\underline{r}^{\prime}\right|}-\delta\rho(\underline{r})\label{Koperator}.
\end{eqnarray}
The integral in the above expression is carried out over the domain
where $\rho_0 > 0$, that is, the general ellipsoidal domain with
radii $R_x, R_y, R_z$ of the unperturbed condensate. Extending the
integration domain to the region where $\rho_0 + \delta\rho > 0$
adds higher order effects since it is exactly in this domain that
$\rho_0 = \mathcal{O}(\delta \rho)$. To investigate the stability of
the BEC we look for eigenfunctions and eigenvalues of the operator
(\ref{stability}): dynamical instability arises when one or more
eigenvalues $\lambda$ possess a positive real part. 
The size of the real eigenvalues dictates the rate at which the
instability grows. Note that the imaginary eigenvalues of
Eq.~(\ref{stability}) relate to stable collective modes of the
system \cite{Castin01}, e.g. sloshing and breathing, and have been
analysed elsewhere for dipolar BECs \cite{Rick08}. In order to find
such eigenfunctions we follow Refs. \cite{Bijnen07,Sinha01} and
consider a polynomial ansatz for the perturbations in the
coordinates $x,y$ and $z$, of total degree $N$. All operators in
(\ref{stability}), acting on polynomials of degree $N$, result in
polynomials of (at most) the same degree, including the operator $K$.
This latter fact was known to 19th century astrophysicists who
calculated the gravitational potential of a heterogeneous ellipsoid
with polynomial density \cite{Ferrers, Dyson}. The integral
appearing in Eq.\ (\ref{Koperator}) is exactly equivalent to such a
potential. A more recent paper by Levin and Muratov summarises these
results and presents a more manageable expression for the resulting
potential \cite{LevinMuratov}. Hence, using these results the
operator $K$ can be evaluated for a general polynomial density
perturbation $\delta\rho = x^py^qz^r$, with $p,q$ and $r$ being
non-negative integers and $p+q+r \le N$. Therefore, the perturbation evolution
operator (\ref{stability}) can be rewritten as a scalar matrix
operator, acting on vectors of polynomial coefficients, for which
finding eigenvectors and eigenvalues is a trivial computational
task.
\begin{figure}
\centering
\includegraphics[width=7.5cm]{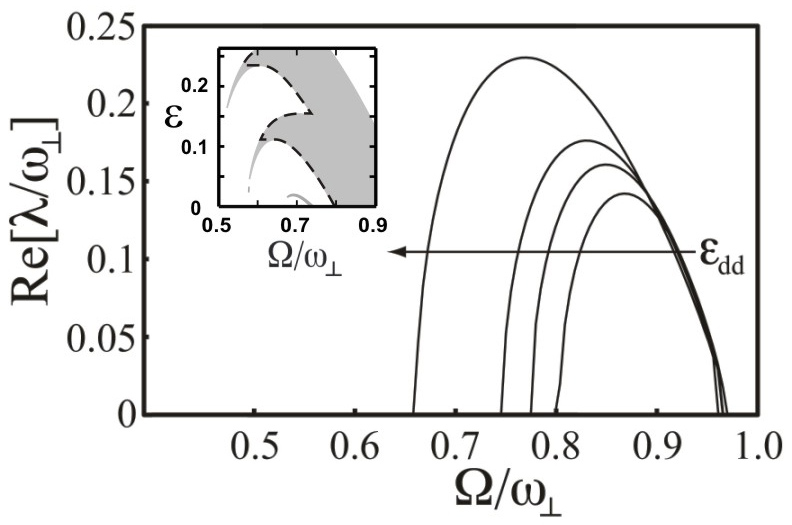}
\caption{The maximum positive real eigenvalues of Eq.\
(\ref{stability}) (solid curves) for the upper-branch solutions of
$\alpha$ as a function of $\Omega$. We assume $\epsilon=0.025$,
$\gamma=1$ and $N=3$, and present various dipolar strengths
$\varepsilon_{dd}=-0.49$, $0$, $0.5$ and $0.99$, with
$\varepsilon_{dd}$ increasing in the direction of the arrow.
The inset shows the full region of dynamical
instability in the $\epsilon-\Omega$ plane for $\varepsilon_{dd}=0$.
The narrow regions have negligible effect and so we only consider
the main instability region (bounded by the dashed line). 
\label{FigEigs}}
 \vspace{-0.5cm}
\end{figure}

Using the above approach we determine the real positive eigenvalues
of Eq.~(\ref{stability}) and thereby predict the regions of
dynamical instability of the static solutions. We focus on the case
of an elliptical trap since this is the experimentally relevant
case.  Recall the general form of the branch diagram
for this case, i.e.\ Fig.~\ref{Fig2}(a).  In the $\alpha<0$
half-plane, the static solutions nearest the $\alpha=0$ axis never
become dynamically unstable, except for a small region $\Omega\simeq
\omega_\perp$, due to a centre-of-mass instability of the condensate
\cite{Rosenbusch02}. The other lower branch solutions are always
dynamically unstable and therefore expected to be irrelevant to
experiment. Thus, we only consider dynamical instability for the
upper branch solutions, i.e.\ the branch in the upper half plane (where $\alpha > 0$).
In Fig.\ \ref{FigEigs} we plot the maximum positive real
eigenvalues of the upper branch solutions as a function of $\Omega$
for a fixed ellipticity $\epsilon=0.025$.  The maximum polynomial perturbation was set at $N=3$, since for this ellipticity it was found that higher order perturbations did not alter the region of instability, and such modes are therefore not displayed.

For a given $\varepsilon_{dd}$ and $\gamma$ there
exists a dynamically unstable region in the $\epsilon-\Omega$ plane.
An illustrative example is shown in Fig.~\ref{FigEigs}(inset) for
$\varepsilon_{dd}=0$ and $\gamma=1$. The instability region (shaded)
consists of a series of crescents  \cite{Sinha01}. Each crescent corresponds to a single value of the polynomial degree $N$, where higher values of $N$ add extra crescents from above. At the high frequency end these
crescents merge to form a main region of instability, characterised
by large eigenvalues.  At the low frequency end the crescents become
vanishingly thin and are characterised by very small eigenvalues
which are at least one order of magnitude smaller than in the main
instability region \cite{Corro07}. As such these regions will only
induce instability in the condensate if they are traversed very
slowly.  This was confirmed by numerical simulations in
Ref.~\cite{Corro07} where it was shown that the narrow instability
regions have negligible effect when ramping $\Omega$ at rates
greater than $d\Omega/dt = 2 \times 10^{-4}\omega_\perp^2$.   It is
unlikely that an experiment could be sufficiently long-lived for
these narrow instability regions to play a role.  For this reason we
will subsequently ignore the narrow regions of instability and
define our instability region to be the main region, as bounded by
the dashed line in Fig.~\ref{FigEigs}(inset). For the experimentally relevant trap ellipticities $\epsilon \ltsimeq 0.1$ the unstable region is defined solely by the $N=3$ perturbations.

We define the lower bound
of the instability region to be $\Omega_i$ (this corresponds to the
dashed line in the inset).  This is the key parameter to
characterise the dynamical instability. As $\varepsilon_{dd}$ is
increased $\Omega_i$ decreases and, accordingly, the unstable range
of $\Omega$ widens. Note that the upper bound of the instability
region is defined by the endpoint of the upper branch at $\Omega \simeq \omega_\perp$.


\section{Routes to instability and vortex lattice formation}

\subsection{Procedures to induce instability}

For a non-dipolar BEC the static solutions and their stability in
the rotating frame depend only on rotation frequency $\Omega$ and
trap ellipticity $\epsilon$. Adiabatic changes in $\epsilon$ and
$\Omega$ can be employed to evolve the condensate through the static
solutions and reach a point of instability. Indeed, this has been
realized both experimentally \cite{Madison00,Hodby02} and
numerically \cite{Lundh03,Parker06}, with excellent agreement to the
hydrodynamic predictions. For the case of a dipolar BEC we have
shown in Sections~\ref{SecStationaryStates} and \ref{SecStability}
that the static solutions and their instability depend additionally
on the trap ratio $\gamma$ and the interaction parameter
$\varepsilon_{dd}$. Since all of these parameters can be
experimentally tuned in time, one can realistically consider each
parameter as a distinct route to traverse the parameter space of
solutions and induce instability in the system.

\begin{figure}
\centering
\includegraphics[width=7.5cm]{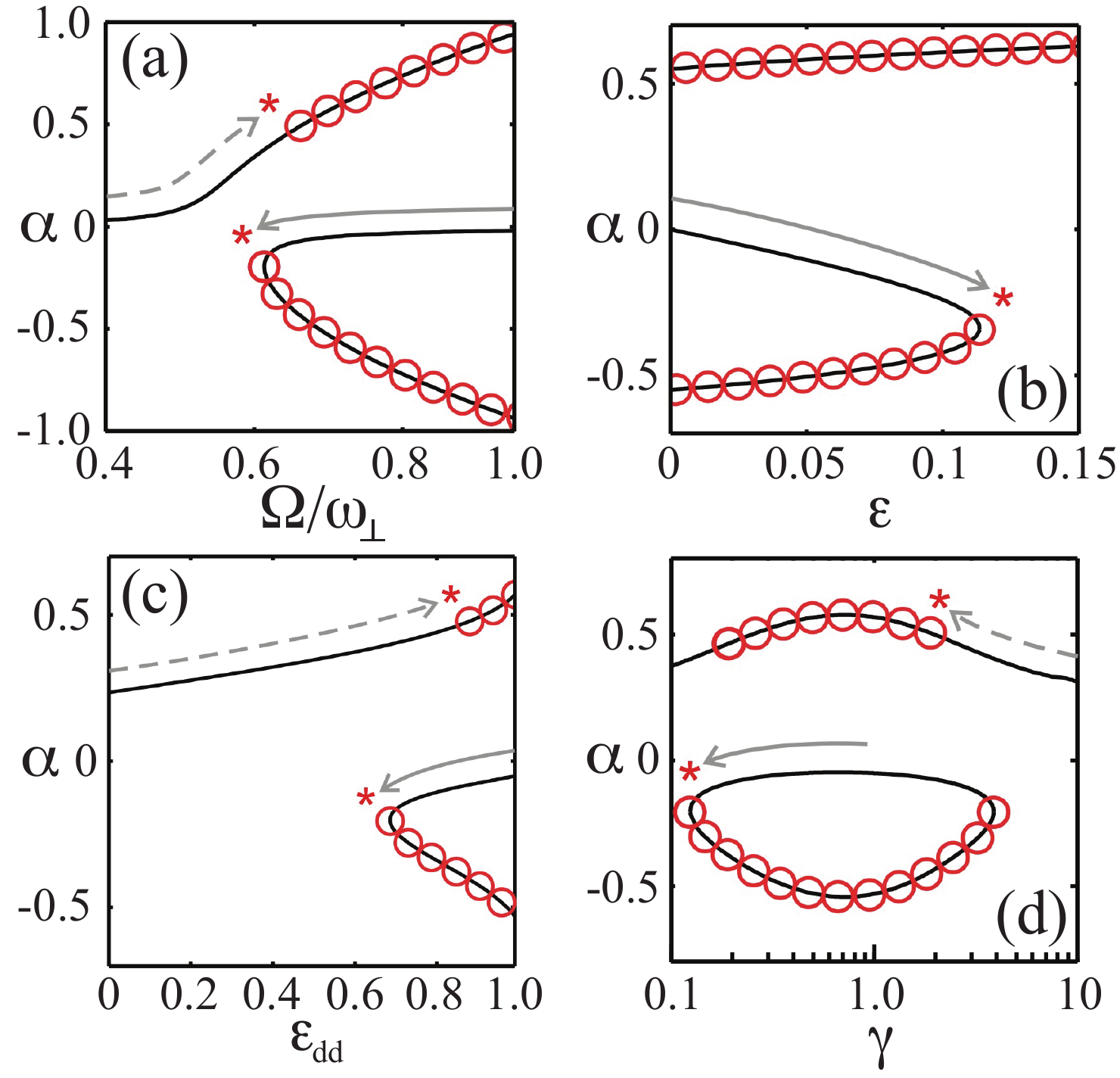}
\caption{Stationary states in the rotating trap characterised by the
velocity field amplitude $\alpha$, determined from Eq.\
(\ref{alpha}). Dynamically unstable solutions are marked with red
circles. In each of the figures the trap rotation frequency $\Omega$
(a), trap ellipticity $\epsilon$ (b), dipolar interaction strength
$\edd$ (c) and axial trapping strength $\gamma$ (d) are varied
adiabatically, whilst the remaining parameters remain fixed at
$\Omega = 0.7 \omega_{\perp}$, $\epsilon = 0.025$, $\edd = 0.99$,
and  $\gamma = 1$. The adiabatic pathways to instability (onset marked by red
asterisk) are schematically shown by the dashed and solid arrows.
Dashed arrows indicate a route towards dynamical instability,
whereas solid arrows indicate an instability due to disappearance of
the stationary state.\label{FigRoutes}}
\end{figure}

Examples of these routes are presented in Fig.~\ref{FigRoutes}.
Specifically, Fig.~\ref{FigRoutes} shows the static solutions
$\alpha$ of Eq.\ (\ref{alpha}) as a function of $\Omega$ [Fig.\ 
\ref{FigRoutes}(a)], $\epsilon$ [Fig.\ \ref{FigRoutes}(b)], $\edd$
[Fig.\ \ref{FigRoutes}(c)] and $\gamma$ [Fig.\ \ref{FigRoutes}(d)]. In
each case the remaining three parameters are fixed at $\epsilon =
0.025$, $\gamma = 1$, $\Omega = 0.7\omega_{\perp}$, and $\edd =
0.99$. Dynamically unstable solutions are indicated with red
circles. Grey arrows mark routes towards instability (the point of onset of instability being marked by an asterisk),
where the free parameter $\Omega$, $\epsilon$, $\edd$, or $\gamma$
is varied adiabatically until either a dynamical instability is
reached, or the solution branch backbends and so ceases to exist. For
solutions with $\alpha >0$, the instability is always due to the
system becoming dynamically unstable (dashed arrows), whereas for
$\alpha < 0$ the instability is always due to the solution branch
backbending on itself (solid arrows) and so ceasing to exist. Numerical studies
\cite{Parker06} indicate that these two types of instability involve
different dynamics and possibly have distinct experimental
signatures.

Below we describe adiabatic variation of each parameter in more
general detail, beginning with the established routes towards
instability in which (i) $\Omega$ and (ii) $\epsilon$ are varied,
and then novel routes based on adiabatic changes in (iii)
$\varepsilon_{dd}$ and (iv) $\gamma$. In each case it is crucial to
consider the behaviour of the points of instability, namely the
back-bending point $\Omega_b$ and the onset of dynamical instability
of the upper branch $\Omega_i$.
%
%
\begin{figure}
\centering
\includegraphics[width=7.5cm]{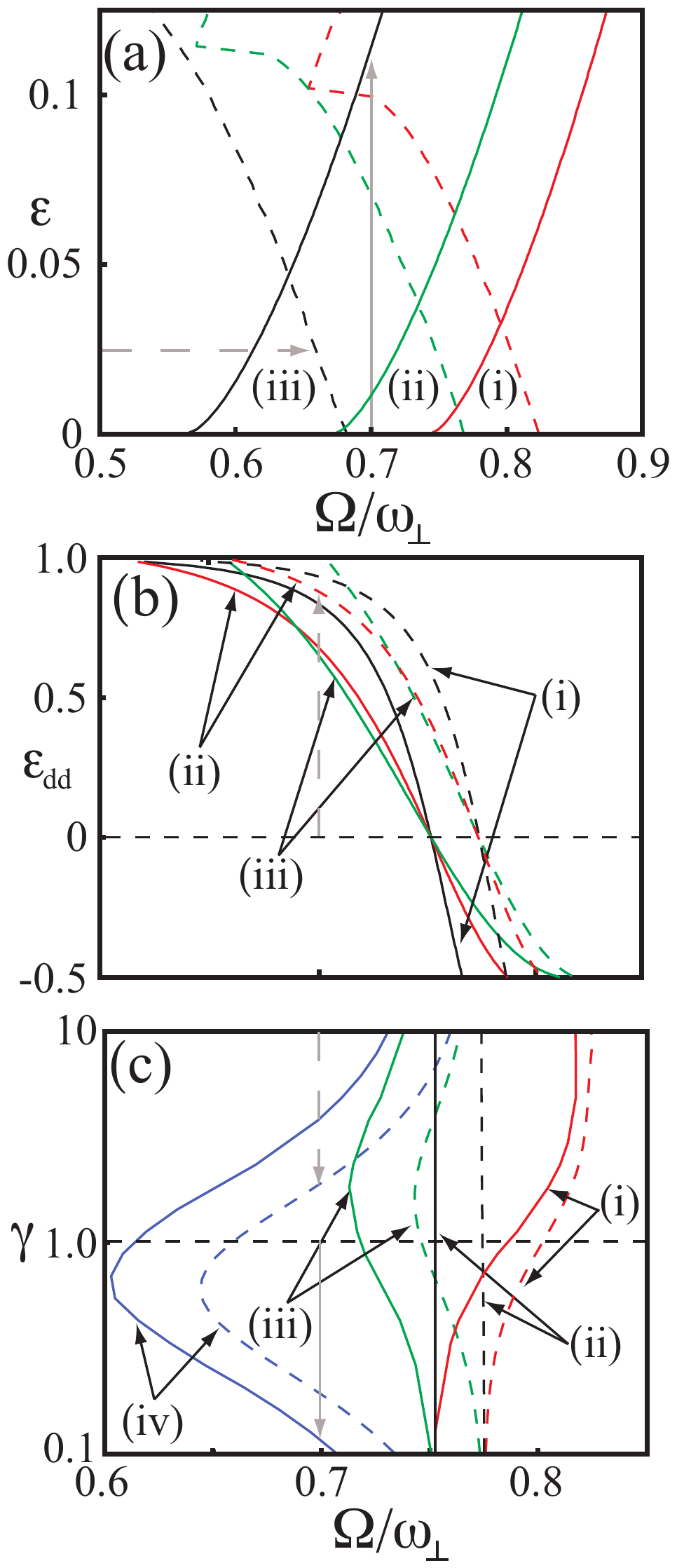}
\caption{(a)
Phase diagram of $\epsilon$ versus $\Omega_b$ (solid curves) and
$\Omega_i$ (dashed curves) for $\gamma=1$ and (i)
$\varepsilon_{dd}=-0.49$, (ii) $0.5$ and (iii) $0.99$. (b) Phase diagram of $\varepsilon_{dd}$ versus $\Omega_b$
(solid curves) and $\Omega_i$ (dashed curves) for $\epsilon=0.025$
and (i) $\gamma=0.5$, (ii) $1$ and (iii) $2$. (c) Phase diagram of
$\gamma$ versus $\Omega_b$ (solid curves) and $\Omega_i$ (dashed
curves) for $\epsilon=0.025$ and (i) $\varepsilon_{dd}=-0.49$, (ii)
$0$, (iii) $0.5$ and (iv) $0.99$. In each case the solid (dashed) arrows depict the routes to instability shown in Fig.\ 4.} \label{FigObOi}
 \vspace{-0.5cm}
\end{figure}

\textbf{i) Adiabatic introduction of $\Omega$:} The relevant
parameter space of $\epsilon$ and $\Omega$ is presented in
Fig.~\ref{FigObOi}(a), with the instability frequencies
$\Omega_b(\epsilon)$ (solid curves) and $\Omega_i(\epsilon)$ (dashed
curves) indicated. For a BEC initially confined to a non-rotating
trap with finite ellipticity $\epsilon$, as the rotation frequency
$\Omega$ is increased adiabatically  the BEC follows the upper
branch solution [Fig.\ \ref{FigRoutes}(a) (dashed arrow)]. This
particular route traces out a horizontal path in Fig.\
\ref{FigObOi}(a) until it reaches $\Omega_i(\epsilon)$, where the
stationary solution becomes dynamically unstable. For the specific parameters of
Fig.~\ref{FigRoutes}(a) the system becomes unstable at
$\Omega=\Omega_i(\epsilon)\approx 0.65 \omega_\perp$.   More
generally, Fig.~\ref{FigObOi}(a) shows that as $\varepsilon_{dd}$
is increased, $\Omega_i(\epsilon)$ is decreased and as such
instabilities in the stationary solutions will occur at lower
rotation frequencies. At $\epsilon \simeq 0.1$, the curve for
$\Omega_i$ displays a sharp kink, arising from the shape of the dynamically unstable region, as shown in
Fig.~\ref{FigEigs}(inset).

\textbf{ii) Adiabatic introduction of $\epsilon$:} Here we begin
with a cylindrically symmetric ($\epsilon=0$) trap, rotating at a
fixed frequency $\Omega$. The trap ellipticity $\epsilon$ is then
increased adiabatically and in the phase diagram of Fig.\
\ref{FigObOi}(a) the BEC traces out a vertical path starting at
$\epsilon = 0$. The ensuing dynamics depend on the trap rotation
speed relative to $\Omega_b(\epsilon=0)$:

(a) For $\Omega <
\Omega_b(\epsilon=0)$ the condensate follows the upper branch of the
static solutions shown in Fig.\ \ref{FigRoutes}(a). This branch moves progressively to larger $\alpha$. For
$\Omega<\Omega_i(\epsilon)$ the BEC remains stable but as $\epsilon$
is increased further the condensate eventually becomes dynamically
unstable. Figure \ref{FigObOi}(a) shows that as $\varepsilon_{dd}$
is increased $\Omega_i(\epsilon)$ is decreased and as such the
dynamical instability of the stationary solutions occurs at a lower
trap ellipticity.

(b) For $\Omega > \Omega_b(\epsilon=0)$ the condensate accesses the
lower branch solutions nearest the $\alpha=0$ axis. These solutions
are always dynamically stable and the criteria for instability is
instead determined by whether the solution exists. As $\epsilon$ is
increased the back-bending frequency $\Omega_b(\epsilon)$ increases.
Therefore, when $\epsilon$ exceeds some critical value the lower
branch solutions disappear for the chosen value of rotation
frequency $\Omega$. This occurs when $\Omega<\Omega_b(\epsilon)$.
Figure \ref{FigObOi}(a) shows that as $\varepsilon_{dd}$ is
increased $\Omega_b(\epsilon)$ is decreased and as such
instabilities in the system will occur at a higher trap ellipticity.
At this point the parabolic condensate density profile no longer represents a stable solution.
The particular route indicated in Fig.\ \ref{FigRoutes}(b) is
included in Fig.\ \ref{FigObOi}(a) as a vertical, solid grey arrow.

\textbf{iii) Adiabatic change of $\varepsilon_{dd}$:} The relevant
parameter space of $\varepsilon_{dd}$ and $\Omega$ is shown in
Fig.~5(b) for several different trap ratios.  Consider that we begin
from an initial BEC in a trap with finite ellipticity
$\epsilon=0.025$ and rotation frequency $\Omega$.  (This can be
achieved, for example, by increasing $\epsilon$ from zero at fixed
$\Omega$.) Then, by changing $\varepsilon_{dd}$ adiabatically an
instability can be induced in two ways:

(a) For $\Omega < \Omega_b(\varepsilon_{dd})$ the condensate follows
the upper branch solutions until they become unstable. This route to
instability in shown in Fig.~4(c) by the dashed arrow, with the
corresponding path in Fig.~5(b) shown by the vertical dashed arrow.
Thus for $\Omega < \Omega_i(\varepsilon_{dd})$ the motion remains
stable. However, for $\Omega > \Omega_i(\varepsilon_{dd})$ the upper
branch becomes dynamically unstable. In Fig.\ \ref{FigObOi}(b)
$\Omega_i(\varepsilon_{dd})$ (dashed curves) is plotted for
different trap ratios. As can be seen, the stable region of the upper branch becomes smaller as $\varepsilon_{dd}$ is increased.

(b) For $\Omega > \Omega_b(\varepsilon_{dd})$ the condensate follows
the lower branch solutions nearest the $\alpha=0$ axis. These
solutions are always stable and hence an instability can only be
induced when this solution no longer exists, i.e. $\Omega <
\Omega_b(\varepsilon_{dd})$. Figure \ref{FigObOi}(b) shows
$\Omega_b(\varepsilon_{dd})$ (solid curves) for various trap aspect
ratios. As can be seen the back-bending frequency $\Omega_b$
decreases as $\varepsilon_{dd}$ is increased. Thus if
$\varepsilon_{dd}$ is increased the system will remain stable.
However if $\varepsilon_{dd}$ is decreased then the system will
become unstable when $\Omega =\Omega_b(\varepsilon_{dd})$.

\textbf{iv) Adiabatic change of $\gamma$:} Figure \ref{FigObOi}(c) shows the
parameter space of $\gamma$ and $\Omega$. Consider, again, an
initial stable condensate with finite trap rotation frequency
$\Omega$ and ellipticity $\epsilon=0.025$. Then through adiabatic
changes in $\gamma$ the condensate can traverse the parameter space
and, depending on the initial conditions, the instability can arise
in two ways:

(a) For $\Omega < \Omega_b(\gamma)$ the condensate exists on the
upper branch. It is then relevant to consider the onset of dynamical
instability $\Omega_i(\gamma)$ (dashed curves in
Fig.~\ref{FigObOi}(c)). Providing $\Omega < \Omega_i(\gamma)$ the
solution remains dynamically stable. However, once
$\Omega>\Omega_i(\gamma)$ the upper branch solutions become
unstable.

(b) For $\Omega > \Omega_b(\gamma)$ the condensate exists on the
lower branch nearest the $\alpha=0$ axis. These solutions are always
dynamically stable and instability can only occur when the motion of
the back-bending point causes the solution to disappear. This occurs
when $\Omega < \Omega_b(\gamma)$, with $\Omega_b(\gamma)$ shown in
Fig.~\ref{FigObOi}(c) by solid curves for various dipolar interaction strengths.

These two paths to instability are shown in Fig.\ \ref{FigRoutes}(d)
and are also indicated in Fig.\ \ref{FigObOi}(c) as vertical grey
arrows, where the dashed (solid) arrow corresponds to the $\alpha >
0$ ($\alpha < 0$) path.

\subsection{Is the final state of the system a vortex lattice?}

Having revealed the points at which a rotating dipolar condensate
becomes unstable we will now address the question of whether this
instability leads to a vortex lattice. First, let us review the
situation for a non-dipolar BEC. The presence of vortices in the
system becomes energetically favorable when the rotation frequency
exceeds a critical frequency $\Omega_v$. Working in the TF limit,
with the background density taking the parabolic form (\ref{TF}),
$\Omega_v$ can be approximated as \cite{lundh97},
\begin{equation}
\Omega_v=\frac{5}{2}\frac{\hbar}{mR^2}\ln \frac{0.67R}{\xi_s}.
\label{crit_freq}
\end{equation}
Here the condensate is assumed to be circularly symmetric with
radius $R$, and $\xi_s=\hbar/\sqrt{2m \rho_0 g}$ is the healing
length that characterises the size of the vortex core. For typical
condensate parameters $\Omega_v \sim 0.4 \omega_\perp$. It is
observed experimentally, however, that vortex lattice formation
occurs at considerably higher frequencies, typically $\Omega \sim 0.7
\omega_\perp$. This difference arises because above $\Omega_v$ the
vortex-free solutions remain remarkably stable. It is only once a
hydrodynamic instability occurs (which occurs in the locality of
$\Omega \approx 0.7\omega_\perp$) that the condensate has a
mechanism to deviate from the vortex-free solution and relax into a
vortex lattice. Another way of visualising this is as follows. Above
$\Omega_v$ the vortex-free condensate resides in some local energy
minimum, while the global minimum represents a vortex or vortex
lattice state. Since the vortex is a topological defect, there
typically exists a considerable energy barrier for a vortex to enter
the system. However, the hydrodynamic instabilities offer a route to
navigate the BEC out of the vortex-free local energy minimum towards
the vortex lattice state.
\begin{figure}
\centering
\includegraphics[width=7.5cm]{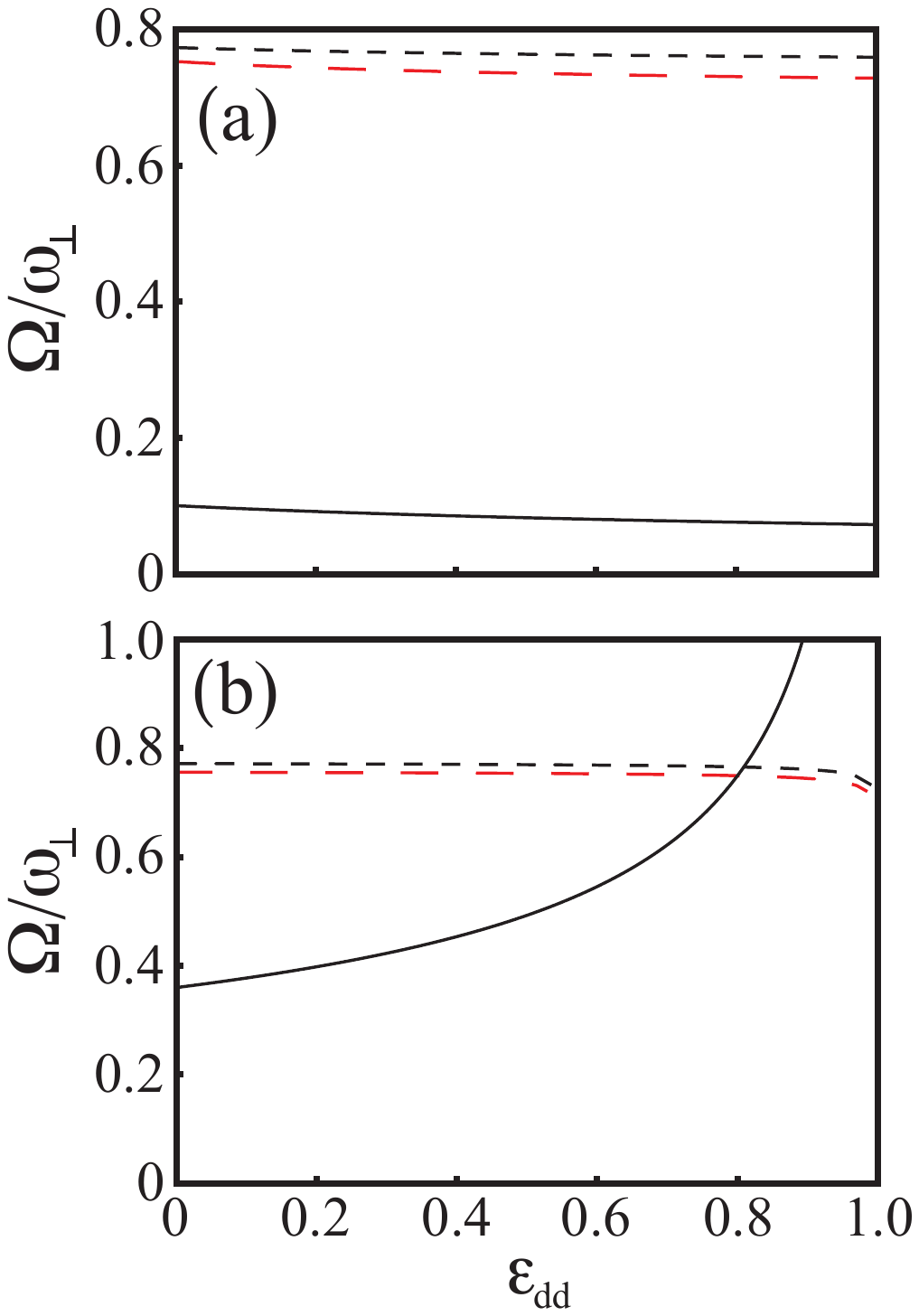}
\caption{The relation between the instability frequencies,
$\Omega_b$ (long dashed red curve) and $\Omega_i$ (short dashed
curve), and the critical rotation frequency for vorticity $\Omega_v$
(solid curve) for (a) an oblate trap $\gamma=10$ and (b) a prolate
trap $\gamma=0.1$. The instability frequencies are based on a trap
with ellipticity $\epsilon=0.025$ while $\Omega_v$ is obtained from
Eq.~(\ref{crit_freq}) under the assumption of a $^{52}$Cr BEC with
$150,000$ atoms and scattering length $a_s=5.1$nm in a circularly
symmetric trap with $\omega_\perp=2\pi \times 200$Hz.\label{Fig5}}
 \vspace{-0.5cm}
\end{figure}

Note that vortex lattice formation occurs via non-trivial dynamics.
The initial hydrodynamic instability in the vortex-free state that
we have discussed in this paper is only the first step
\cite{Parker06}. For example, if the condensate is on the upper
branch of hydrodynamic solutions (e.g. under adiabatic introduction
of $\Omega$) and undergoes a dynamical instability, this leads to
the exponential growth of surface ripples in the condensate
\cite{Parker06,Madison00}. Alternatively, if the condensate is on
the lower branch and the static solutions disappear (e.g. following
the introduction of $\epsilon$) the condensate undergoes large and
dramatic shape oscillations. In both cases the destabilisation of
the vortex-free condensate leads to the nucleation of vortices into
the system. A transient turbulent state of vortices and density
perturbations then forms, which subsequently relaxes into a vortex
lattice configuration \cite{Parker06,Parker05}.

In the presence of dipolar interactions, however, the critical
frequency for a vortex depends crucially on the trap geometry
$\gamma$ and the strength of the dipolar interactions
$\varepsilon_{dd}$. Following reference \cite{ODell07} we will make
a simple and approximated extension of Eq.~(\ref{crit_freq}) to a
dipolar BEC. We will consider a circularly-symmetric dipolar
condensate with radius $R=R_x=R_y$ that satisfies
Eqs.~(\ref{kx})-(\ref{size}), and insert this into
Eq.~(\ref{crit_freq}) for the condensate radius. This method still
assumes that the size of the vortex is characterised by the s-wave
healing length $\xi_s$. Although one does expect the dipolar
interactions to modify the size of the vortex core, it should be noted that
Eq.~(\ref{crit_freq}) only has logarithmic accuracy and is
relatively insensitive to the choice of vortex core length scale. The
dominant effect of the dipolar interactions in Eq.~(\ref{crit_freq})
comes from the radial size and is accounted for. Note also that this
expression is for a circularly symmetric system while we are largely
concerned with elliptical traps. However we will employ a very weak
ellipticity $\epsilon=0.025$ for which we expect the correction to
the critical frequency to be correspondingly small.

As an example, we take the parameter space of rotation
frequency $\Omega$ and dipolar interactions $\varepsilon_{dd}$. We
first consider the behavior in a quite oblate trap with
$\gamma=10$. In Fig.~\ref{Fig5}(a) we plot the instability frequencies
$\Omega_i$ and $\Omega_b$ for this system as a function of the
dipolar interactions $\varepsilon_{dd}$. Depending on the specifics
of how this parameter space is traversed, either by adiabatic
changes in $\Omega$ (vertical path) or $\varepsilon_{dd}$
(horizontal path), the condensate will become unstable when it
reaches one of the instability lines (short and long dashed lines).
These points of instability decrease weakly with dipolar
interactions and have the approximate value $\Omega_i\approx
\Omega_b \approx 0.75\omega_\perp$. On the same plot we present the
critical rotation frequency $\Omega_v$ according to
Eq.~(\ref{crit_freq}). In order to calculate this we have assumed a
BEC of 150,000 $^{52}$Cr atoms confined within a trap with
$\omega_\perp=2\pi \times 200$Hz. In this oblate system we see that
the dipolar interactions lead to a decrease in $\Omega_v$, as noted
in \cite{ODell07}. This dependence is very weak at this value of $\gamma$, and throughout the
range of $\varepsilon_{dd}$ presented it maintains the approximate
value $\Omega_v \approx0.1\omega_{\perp}$. Importantly these results show that
when the condensate becomes unstable a vortex/vortex lattice
state is energetically favored. As such, we expect that in an
oblate dipolar BEC a vortex lattice will ultimately form when these
instabilities are reached.

In Fig.~\ref{Fig5}(b), we make a similar plot but for a prolate trap with
$\gamma=0.1$. The instability frequencies show a somewhat similar
behaviour to the oblate case. However, $\Omega_v$ is drastically
different, increasing significantly with $\varepsilon_{dd}$. We
find that this qualitative behaviour occurs consistently in prolate
systems, as noted in \cite{ODell07}. This introduces two regimes
depending on the dipolar interactions. For $\varepsilon_{dd}
\ltsimeq 0.8$, $\Omega_{i,b}>\Omega_v$, and so we expect a
vortex/vortex lattice state to form following the instability.
However, for $\varepsilon_{dd} \gtsimeq 0.8$ we find an intriguing
new regime in which $\Omega_{i,b}<\Omega_v$. In other words, while the
instability in the vortex-free parabolic density profile still occurs, a vortex state is not energetically favorable. The
final state of the system is therefore not clear. Given that a
prolate dipolar BEC is dominated by attractive interactions (since
the dipoles lie predominantly in an attractive end-to-end
configuration) one might expect similar behavior to the case of
conventional BECs with attractive interactions ($g<0$) where the
formation of a vortex lattice can also be energetically unfavorable. Suggestions for
final state of the condensate in this case include centre-of-mass motion and collective oscillations, such as quadrupole modes or higher angular momentum-carrying shape excitations \cite{Wilkin98,Mottelson99,Pethick00}.
However the nature of the true final state in this case is beyond the scope
of this work and warrants further investigation.



\section{Conclusions}

By calculating the static hydrodynamic solutions of a rotating
dipolar BEC and studying their stability, we have predicted the
regimes of stable and unstable motion. In general we find that the
backbending or bifurcation frequency $\Omega_b$ decreases with increasing dipolar
interactions. In addition, the onset of dynamical instability in the
upper branch solutions, $\Omega_i$, decreases with increasing
dipolar interactions. Furthermore these frequencies depend on the
aspect ratio of the trap.

By utilising the novel features of dipolar condensates we detail
several routes to traverse the parameter space of static solutions
and reach a point of instability. This can be achieved through
adiabatic changes in trap rotation frequency $\Omega$, trap
ellipticity $\epsilon$, dipolar interactions $\varepsilon_{dd}$ and
trap aspect ratio $\gamma$, all of which are experimentally tunable
quantities. While the former two methods have been employed for
non-dipolar BECs, the latter two methods are unique to dipolar BECs.
In an experiment the latter instabilities would therefore demonstrate the special role played by dipolar interactions.
Furthermore, unlike for conventional BECs with repulsive interactions,
the formation of a vortex lattice following a hydrodynamic instability is
not always favored and depends sensitively on the shape of the
system. For a prolate BEC with strong dipolar interactions there
exists a regime in which the rotating spheroidal parabolic Thomas-Fermi density profile is unstable and yet it
is energetically unfavorable to form a lattice. Other outcomes may
then develop, such as a centre-of-mass motion of the system or
collective modes with angular momentum. However, for oblate dipolar
condensates, as well as prolate condensates with weak dipolar
interactions, the presence of vortices is energetically favored at
the point of instability and we expect the instability to lead to
the formation of a vortex lattice.

We acknowledge financial support from the Australian Research
Council (AMM), the Government of Canada (NGP) and the Natural
Sciences and Engineering Research Council of Canada (DHJOD).


\begin{thebibliography}{99}
\bibitem{Griesmaier05} A. Griesmaier,J. Werner, S. Hensler, J. Stuhler and T. Pfau, Phys. Rev. Lett. {\bf 94}, 160401 (2005).
\bibitem{Lahaye07}T. Lahaye, T. Koch, B. Frohlich, M. Fattori, J. Metz, A. Griesmaier, S. Giovanazzi and T. Pfau, Nature (London) {\bf 448}, 672 (2007).
\bibitem{Koch08}T. Koch, T. Lahaye, J. Metz, B. Frohlich, A. Griesmaier and T. Pfau, Nat. Phys. {\bf 4}, 218 (2008).
\bibitem{Santos00} L. Santos, G.V. Shlyapnikov, P. Zoller, and M. Lewenstein, Phys.
Rev. Lett. \textbf{85}, 1791 (2000).
\bibitem{Yi01} S. Yi and L. You, Phys. Rev. A {\bf 63}, 053607 (2001).
\bibitem{Pethick&Smithbook} C. J. Pethick and H. Smith,
\textit{Bose-Einstein Condensation in Dilute Gases} (Cambridge
University Press, 2002).
\bibitem{Pitaevskii&StringariBook}
L. Pitaevskii and S. Stringari, \textit{Bose-Einstein Condensation}
(Oxford, 2003), p183.
\bibitem{Pines&NozieresBook}
P. Nozieres and D. Pines, \textit{Theory Of Quantum Liquids} 
(Westview, New York, 1999).
\bibitem{stringari96}
{S. Stringari, Phys. Rev. Lett. \textbf{77}, 2360 (1996).}
\bibitem{Parker08} N.G. Parker and D. H. J.
O'Dell, Phys. Rev. A {\bf 78}, 041601(R) (2008).
\bibitem{ODell04} D. H. J. O'Dell, S. Giovanazzi and C. Eberlein,
Phys. Rev. Lett. {\bf 92}, 250401 (2004).
\bibitem{Eberlein05} C. Eberlein, S. Giovanazzi and D.H.J. O'Dell, Phys. Rev. A {\bf 71}, 033618 (2005).

\bibitem{Madison00} K. W. Madison, F. Chevy, W. Wohlleben and J. Dalibard, Phys. Rev. Lett. {\bf 84}, 806, (2000); K. W. Madison, F. Chevy, V. Bretin and J. Dalibard, {\it ibid.} {\bf 86} 4443 (2001).
\bibitem{Hodby02} E. Hodby {\it et al.}, Phys. Rev. Lett. {\bf 88}, 010405 (2002).
\bibitem{Recati01} A. Recati, F. Zambelli and S. Stringari, Phys. Rev. Lett. {\bf 86}, 377 (2001).
\bibitem{Sinha01} S. Sinha and Y. Castin, Phys. Rev. Lett. {\bf 87}, 190402 (2001).
\bibitem{Lundh03} E. Lundh, J.-P. Martikainen and K.-A. Suominen, Phys. Rev. A {\bf 67}, 063604 (2003).
\bibitem{Parker06} N. G. Parker, R. M. W. van Bijnen and A. M. Martin, Phys. Rev. A {\bf 73}, 061603(R) (2006).
\bibitem{Corro07} I. Corro, N. G. Parker and A. M. Martin, J. Phys. B {\bf 40}, 3615
(2007).


\bibitem{Pu06} S. Yi and H. Pu, Phys. Rev. A {\bf 73}, 061602(R) (2006).
\bibitem{ODell07} D. H. J. O'Dell and C. Eberlein, Phys. Rev. A {\bf 75}, 013604 (2007).

\bibitem{Wilson08} R. M. Wilson, S. Ronen, J. L. Bohn, and H. Pu, Phys. Rev. Lett. 100, 245302 (2008).
\bibitem{Cooper05} N. R. Cooper, E. H. Rezayi and S. H. Simon, Phys. Rev. Lett. {\bf 95} 200402 (2005).
\bibitem{Zhang05} J. Zhang and H. Zhai, Phys. Rev. Lett. {\bf 95}, 200403 (2005).
\bibitem{Cooper07} S. Komineas and N. R. Cooper, Phys. Rev. A {\bf 75}, 023623 (2007).





\bibitem{Bijnen07} R. M. W. van Bijnen, D. H. J. O'Dell, N. G. Parker and A. M. Martin, Phys. Rev. Lett.
{\bf 98}, 150401 (2007).
\bibitem{Martin08} A. M. Martin, N. G. Parker, R. M. W. van Bijnen, A. Dow and D. H. J. O'Dell, Las. Phys. {\bf 18}, 322 (2008).
\bibitem{Goral00} K. G{\'{o}}ral, K. Rz{\c{a}}{\.{z}}ewski, and T. Pfau, Phys. Rev. A {\bf 61}, 051601 (2000).
\bibitem{Yi00} S. Yi and L. You, Phys. Rev. A {\bf 61}, 041604 (2000).
\bibitem{You98} M. Marinescu and L. You, Phys. Rev. Lett. {\bf 81}, 4596 (1998).

\bibitem{Giovanazzi02}S. Giovanazzi, A. Gorlitz and T. Pfau, Phys. Rev. Lett. {\bf 89}, 130401 (2002).
\bibitem{Parker09} N. G. Parker, C. Ticknor, A. M. Martin and D. H.
O'Dell, Phys. Rev. A {\bf 79}, 013617 (2009).

\bibitem{Leggett_Book} A. J. Leggett, \textit{Quantum Liquids: Bose-Condensation and Cooper Pairing in Condensed Matter Systems} (Oxford University Press, 2006).
\bibitem{Leggett00} A. J. Leggett, in {\it Bose-Einstein Condensation: From Atomic Physics to Quantum Fluids}, Proceedings of the Thirteenth Physics Summer School, edited by C. M. Savage and M. P. Das, (World Scientific, 2001), p. 1-42.
\bibitem{Lifshitz} L. D. Landau and E. M. Lifshitz, {\it Course of Theoretical Physics: Mechanics} (Butterworth-Heinemann, third edition, 1982)

\bibitem{Gradshteyn00} L. S. Gradshteyn and I. M. Ryzhik, eds., {\it Table of Integrals, Series, and Products} (Academic Press, San Diego, 2000), sixth ed.
\bibitem{Abramowitz74} M. Abramowitz and I. Stegun, eds., {\it Handbook of Mathematical Functions} (Dover, New York, 1974).

\bibitem{Castin01} Y. Castin, in {\it Coherent Matter Waves}, Lecture Notes of Les Houches Summer School, edited by R. Kaiser, C. Westbrook and F. David (Springer-Verlag, 2001), p. 1-136.
\bibitem{Rick08} R. M. W. van Bijnen, N. G. Parker, A. M. Martin and D. H. J. O'Dell, preprint (2008).
\bibitem{Ferrers} N.~M. Ferrers, Quart. J. Pure and Appl. Math {\bf 14}, 1 (1877).
\bibitem{Dyson} F.~W. Dyson, Quart. J. Pure and Appl. Math, {\bf 25}, 259 (1891).
\bibitem{LevinMuratov} M.~L. Levin and R.~Z. Muratov, Astrophys. J. {\bf 166},
441 (1971).

\bibitem{Rosenbusch02} P. Rosenbusch, D.S. Petrov, S. Sinha, F. Chevy, V. Bretin,
Y.  Castin, G. Shlyapnikov, and J. Dalibard, Phys.  Rev.  Lett.  \textbf{88}, 250403,
(2002)

\bibitem{Lundh97} E. Lundh, C.J. Pethick and H. Smith,
Phys. Rev. A {\bf 55}, 2126 (1997).
\bibitem{Wilkin98} N. K. Wilkin, J.M.F. Gunn and R.A. Smith, Phys. Rev. Lett. {\bf 80}, 2265 (1998).
\bibitem{Mottelson99} B. Mottelson, Phys. Rev. Lett. {\bf 83}, 2695 (2000).
\bibitem{Pethick00} C. J. Pethick and L. Pitaevskii, Phys. Rev. A {\bf 62}, 033609 (2000).
\bibitem{Parker05} N. G. Parker and C. S. Adams, Phys. Rev. Lett. {\bf 95}, 145301 (2005); J. Phys. B {\bf 39}, 43 (2006).
\bibitem{lundh97} E. Lundh, C. J. Pethink and H. Smith, Phys. Rev. A
{\bf 55}, 2126 (1997).



\end{thebibliography}
\end{document}